\documentclass[preprint,prl,12pt]{revtex4}%
\usepackage{amsfonts}
\usepackage{amsmath}
\usepackage{amssymb}
\usepackage{graphicx}%
\begin{document}
\title{Stress-aging in the electron-glass }
\author{V. Orlyanchik, and Z. Ovadyahu\vspace{0.2in}}
\affiliation{Racah Institute of Physics, The Hebrew University, Jerusalem 91904,
Israel\vspace{0.5in}\vspace{0.8in}}

\begin{abstract}
A new protocol for an aging experiment is studied in the electron-glass phase
of indium-oxide films. In this protocol, the sample is exposed to a non-ohmic
electric field F for a waiting time t$_{w}$ during which the system attempts
to reach a \textit{steady state} (rather than \textit{relax} towards
\textit{equilibrium}). The relaxation of the excess conductance $\Delta$G
after ohmic conditions are restored exhibit simple aging as long as F is not
too large.%
%TCIMACRO{\TeXButton{rvfill}{\vspace*{\fill}}}%
%BeginExpansion
\vspace*{\fill}%
%EndExpansion


\end{abstract}
\maketitle

Aging is a common phenomenon in non-equilibrium systems \cite{1}. The term
`aging' refers to a continuous change in the properties of the system when it
is maintained in some fixed external conditions (such as temperature,
pressure, etc.) for a waiting-time t$_{w}$. This change may be reflected in
the dynamic response of the system due to an application of a post-aging
disturbance. For example, the viscoelastic response of a polymer to a
mechanical stress will depend on the time t$_{w}$ it was `aged' at a
temperature T prior to applying the stress \cite{1}. Systematic studies of
various glassy systems \cite{2} revealed that aging might manifest itself in
different measurements but all share a common feature: After the external
conditions that affect a certain property P are changed, P relaxes towards its
new equilibrium value in a way that reflects both the time t and the `aging'
time t$_{w}$, namely, P(t)=P(t,t$_{w}$).

A more specific form of aging called `simple-aging' has been recently reported
to occur in several glasses \cite{3,4} where P(t,t$_{w}$) could be described
as a simple master function P(t/t$_{w}$).

The experimental protocol usually employed in aging studies involves
relaxation towards an equilibrium state during t$_{w}$. In this note, we
report on a different protocol where the system is under a constant stress F
and attempts to reach a steady state during t$_{w}$. It turns out that the
relaxation that ensues after the stress is relieved exhibits simple aging as
long as the stress is not too large. The master function P(t/t$_{w}$) is
affected by the particular magnitude of the stress, and above a certain field
the relaxation curves fail to collapse. This is shown to correlate with the
loss of memory in the system.

Our experiments were performed using thin films of crystalline In$_{2}%
$O$_{3-x}$ in the hopping regime. The response P is taken as the conductance
G, and the stress F is the electric field applied along the film. Measurements
were carried out at T=4.11K with the samples immersed in liquid $^{4}$He
inside a storage dewar. This enabled high temperature stability over long
times. A germanium thermometer mounted on the sample stage was used to correct
for residual temperature fluctuations and drift. The conductance of the
samples was measured using a two terminal configuration. For measurements with
F%
%TCIMACRO{\TEXTsymbol{>}}%
%BeginExpansion
$>$%
%EndExpansion
10 V/cm G was measured by a dc technique, biasing the sample with a voltage
source (Keithley's K617) while measuring the resulting current (the voltage
across a 10$^{5}\Omega$ series resistor). This procedure was used during the
stress application. For smaller values of F, we used ac techniques employing a
current pre-amplifier (ITHACO 1211) and a lock-in amplifier (PAR 124). This
was also used to measure the conductance G before each run as well as for the
relaxation after F was reset to the ohmic regime (F typically smaller than 1V/cm).

The steps performed in this series of experiments, and results for a typical
case are illustrated in figure 1. The sample conductance and the accompanying
stress-field F were monitored continuously versus time. Initially, G(t) is
recorded while keeping F=F$_{0}$ chosen to be in the ohmic regime (i.e.,
$\partial$G/$\partial$F%
%TCIMACRO{\TEXTsymbol{\vert}}%
%BeginExpansion
$\vert$%
%EndExpansion
F$_{0}\approx$0) to establish a baseline `equilibrium-G'=G(F$_{0}$,0). Then, F
was switched to F$_{n}\gg$F$_{0}$ which caused an appreciable increase in
conductance (figure 1). Finally, having recorded G(F$_{n}$,t) for a time
t=t$_{w}$, F is switched back to its original value F$_{0}$. This results in
an initial sharp decrease of G followed by a slow relaxation process where the
conductance decreases and asymptotically approaches G(F$_{0}$,0) (the dashed
line in figure 1). The relaxation of the excess conductance $\Delta
$G(t)=G(t$\geq$t$_{w}$)-G(F$_{0}$,0) is plotted in figure 2a where the origin
of the time scale t=0 is the time when F$_{0}$ was re-established. The same
$\Delta$G(t) curves are plotted in figure 2b as function of t/t$_{w}$
illustrating a near-perfect data collapse to a master function $\Delta
$G(t/t$_{w}$). It is emphasized that no free parameters are involved in this
collapse; the \textit{only} step taken to get the master function $\Delta
$G(t/t$_{w}$) is dividing each $\Delta$G(t) curve by its \textit{measured}
t$_{w}$.

The master function that results from the present protocol (to be referred to
as `F-protocol') is quite similar to that of the aging protocol used by Vaknin
et al \cite{3} (`V$_{g}$-protocol'). In both cases, $\Delta$G(t/t$_{w}%
$)$\varpropto$-log(t) for t%
%TCIMACRO{\TEXTsymbol{<}}%
%BeginExpansion
$<$%
%EndExpansion
t$_{w}$ and both show equally good simple aging (compare figure 2b with figure
2 in reference 3). Note that these two protocols are fundamentally different.
The V$_{g}$-protocol conforms to the commonly used procedure where during
t$_{w}$ the system is relaxing as manifested by the fact that $\Delta$G(t) is
logarithmically decreasing function of t. By contrast, the system is excited
during t$_{w}$ in the F-protocol, and the associated $\Delta$G increases
logarithmically with time (inset to figure 1). Note that the V$_{g}$-protocol
is carried out under ohmic conditions throughout the entire process while
strong non-ohmic conditions are used during t$_{w}$ in the F-protocol
\cite{5}. During this time, the electronic system absorbs energy from F
\cite{6} and, as will be shown below, some memory of the system is impaired in
result. It is therefore somewhat surprising that the F-protocol yields as good
simple aging as the V$_{g}$ protocol. In fact, the only feature in the master
function that reflects the difference between the two protocols is the
extrapolated value for t/t$_{w}$ to $\Delta$G(t/t$_{w}$)=0 (c.f., figures 2
and 3). In the V$_{g}$-protocol this happens at t'/t$_{w}\equiv$t* which is
usually =1. This is due to a certain symmetry inherent to this protocol
\cite{7}. When this symmetry is impaired e.g., by using large swings of gate
voltages, this t* becomes larger than unity \cite{7,8}. In the F-protocol t*
is consistently larger than unity and increases systematically with F$_{n}$
reaching a value of $\simeq$10 (inset of figure 3) before the curves fail to
collapse (figure 4). This incidentally means that over the range of fields
where simple aging is observed, the master function carries a memory of
\textit{both} t$_{w}$ and the specific value of F$_{n}$ (namely, the value of
t* for a given sample). The inset to figure 3 may be interpreted as implying
that when F$_{n}\rightarrow0,$ t*$\rightarrow1$, which in other words is just
saying that the sample is under "symmetrical" (i.e., \textit{Ohmic})
conditions both during t$_{w}$ and throughout the subsequent relaxation
process. Obviously, this situation cannot be realized in practice with the F-protocol.

When F$_{n}$ exceeds a certain value the $\Delta$G(t) curves fail to collapse
upon normalization by t$_{w}$ (figure 4). For still higher fields $\Delta$G(t)
becomes independent of t$_{w}$ and assumes the `history-free' law \cite{8}
$\Delta$G(t)$\varpropto-$log(t). This presumably results from the fact that a
large F$_{n}$ has a similar (though not exactly equivalent) effect as that of
raising the system temperature. Above some F$_{n}$, this effective temperature
will bring the system to an ergodic state (above the `glass temperature'), and
the ensuing relaxation upon the switch to F$_{0}$ should be similar to a
quench-cool process \cite{8}. Namely, $\Delta$G(t) should contain no memory of
the past and aging behavior is lost as indeed observed.

The influence of the stress-field, and in particular, its detrimental effect
on the memory of the electron glass, can be monitored in a field-effect
experiment \cite{7}. This was performed using a sample configured as a FET
structure by depositing a gate electrode (Au film) on the backside of the
100$\mu$m glass substrate \cite{9}. The sample was cooled to 4.11K holding its
gate voltage V$_{g}$ at 0V, and was allowed it to relax at this temperature
for $\simeq$12 hours. Then, while monitoring G (using ac techniques), V$_{g}$
was swept to +100V, kept there for 15 seconds after which V$_{g}$ was swept to
--100V. The resulting G(V$_{g}$) curve (figure 5) revealed a memory of the
cool-down-V$_{g}$ in the form of a minimum centered at V$_{g}$=0V. After
allowing the system to relax again under V$_{g}$=0, the procedure was repeated
except that during 10 of the 15 seconds dwell-time at V$_{g}$=+100V, a
non-ohmic dc field F$_{n}$ was applied between the source and drain. The
G(V$_{g}$)%
%TCIMACRO{\TEXTsymbol{\vert}}%
%BeginExpansion
$\vert$%
%EndExpansion
F$_{n}$ traces resulting from this procedure exhibit a \textquotedblleft
memory-cusp\textquotedblright\ that has a progressively smaller magnitude when
F$_{n}$ is increased (c.f., figure 5). This illustrates the memory loss caused
by the stress-field as alluded to above. Moreover, above a threshold F$_{n}$
the anomalous cusp at V$_{g}$=0 completely disappears, and G(V$_{g}$) reflects
just the normal (anti-symmetric) form of the field-effect. It is in this range
of fields that the aging behavior is washed out.

In summary, we have demonstrated that the conductance of an electron glass
carries a memory of a non-ohmic electric field F applied in the past as well
as its duration t$_{w}$. This information is reflected in the relaxation of
the excess conductance $\Delta$G(t) monitored following a switch of F (at t=0)
to its ohmic regime. It was also shown that the non-ohmic fields degrade the
memory in the system and that simple aging is obeyed \textit{only as long as
this memory loss is small}. Our experiments thus illustrate that
'simple-aging' and 'memory' are inter-related properties of the electron glass.

Finally, it is remarkable that simple-aging is observed in many different
systems (electron-glass, spin-glass, polymers, viscous-fluids). That such a
simple scaling scheme should so generally hold is a challenge for theory. This
seems to imply the existence of a common feature, non-specific to the type of
glass being studied \cite{10}. To our knowledge, this common ingredient is yet
to be identified.\smallskip\ 

The authors gratefully acknowledge useful discussions with M. Pollak. This
research was supported by a grant administered by the US-Israel Science
Foundation and a grant administered by the German-Israel Science Foundation.

\begin{description}
\item[Figure captions] 
\end{description}

\begin{enumerate}
\item The sample conductance G versus time during a stress-aging experiment.
F$_{0}$=0.5V/cm is used except during t$_{w}$ where F$_{n}$=95V/cm is
maintained. R$_{\square}$=230M$\Omega$ at T=4.11K. The inset illustrates the
logarithmic law by which G increases under a constant F$_{n}$ (for
t$_{w}\approx$5 days and under F$_{n}$=315V/cm in this example).

\item Relaxation curves of the excess conductance after an excitation by
F$_{n}$=100V/cm for different values of t$_{w}$ (a). Sample with R$_{\square}%
$=57M$\Omega$. The same data as in (a) is plotted in (b) versus t/t$_{w}$. The
dashed line shows the extrapolated value of the logarithmic part of the master
function to $\Delta G$(t/t$_{w}$)=0 to define t*.

\item $\Delta G$(t/t$_{w}$) for three different values of the stress-field
F$_{n}$, measured on the same sample (R$_{\square}$=57M$\Omega$). Each
master-function is labeled by its F$_{n}$ (in units of V/cm). At least three
different t$_{w}$ were used in any such plot with t$_{w}$ ranging between 10
to 1620 seconds. The inset shows t* as a function of the stress-field for this
sample (circles ) and for two other samples (R$_{\square}$=11M$\Omega
$-squares, R$_{\square}$=40M$\Omega$-triangles).

\item $\Delta G$(t/t$_{w}$) for the same sample as in figure 3 (R$_{\square}%
$=57M$\Omega$) while using a sufficiently high stress-field such that simple
aging is no longer obeyed.

\item Field effect $\Delta$G(V$_{g}$) traces measured for the same sample as
in figures 3 (R$_{\square}$=57M$\Omega$) illustrating the `loss of memory' due
to various stress fields. See text for the experimental procedure. The trace
taken with 10$^{-1}$V/cm is the \textquotedblleft
baseline-memory\textquotedblright\ for the series. Note that appreciable
reduction in the anomalous cusp (dip around V$_{g}$=0, c.f., reference 7) for
F$_{n}\geq$400V/cm that coincides with the demise of simple aging in this
sample (c.f., figures 3 and 4).
\end{enumerate}


\begin{thebibliography}{99}
\bibitem {5}The relaxation process $\Delta$G(t$\geq$0) is measured under ohmic
conditions in both protocols.
\bibitem {9}Note that this gate-sample separation is much larger than that
usually employed in MOSFET structures. This prevents screening of the
electron-electron interactions by the gate. The F-protocol results reported
here were independent on whether or not the sample had a gate.
\end{thebibliography}
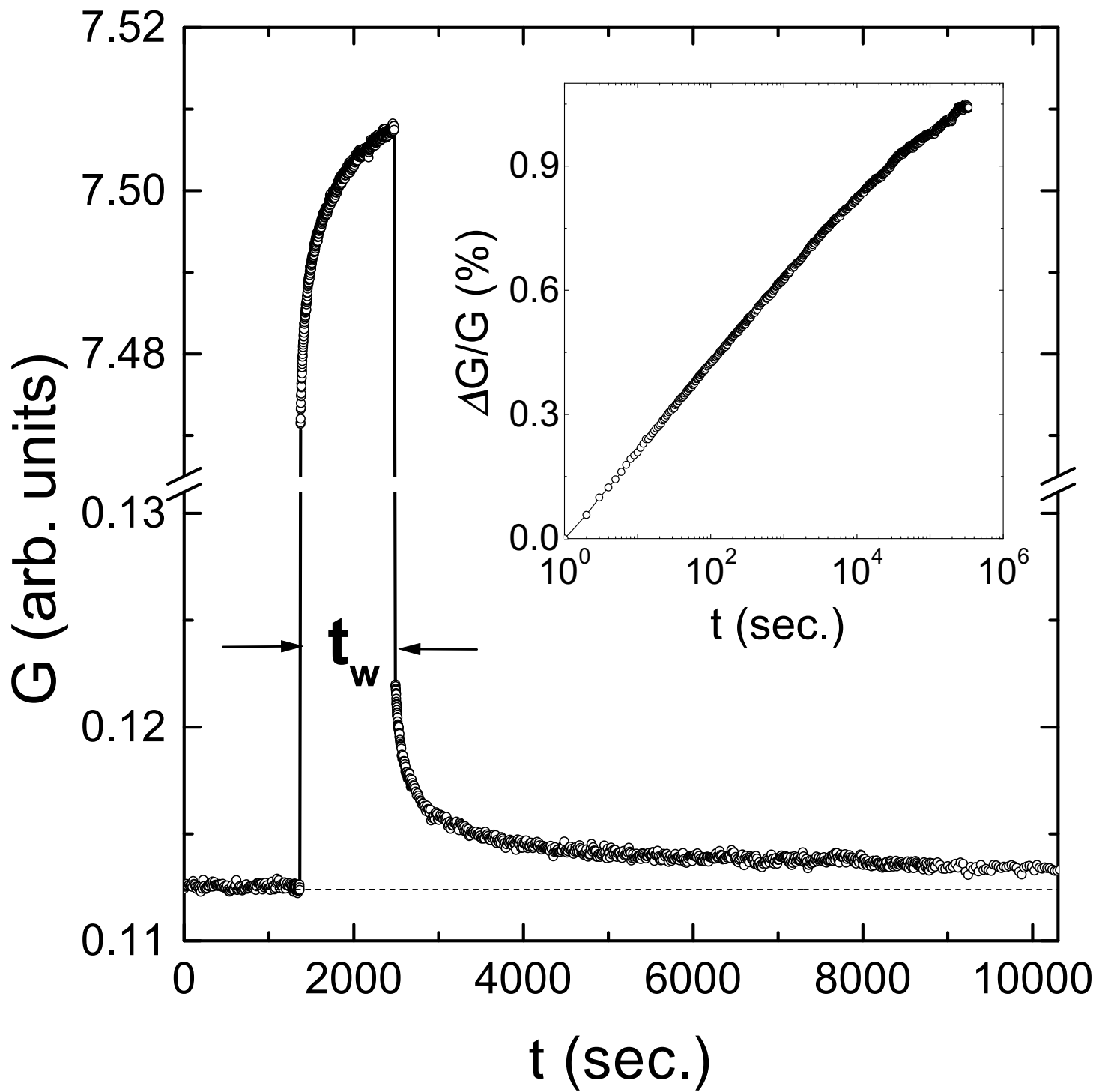
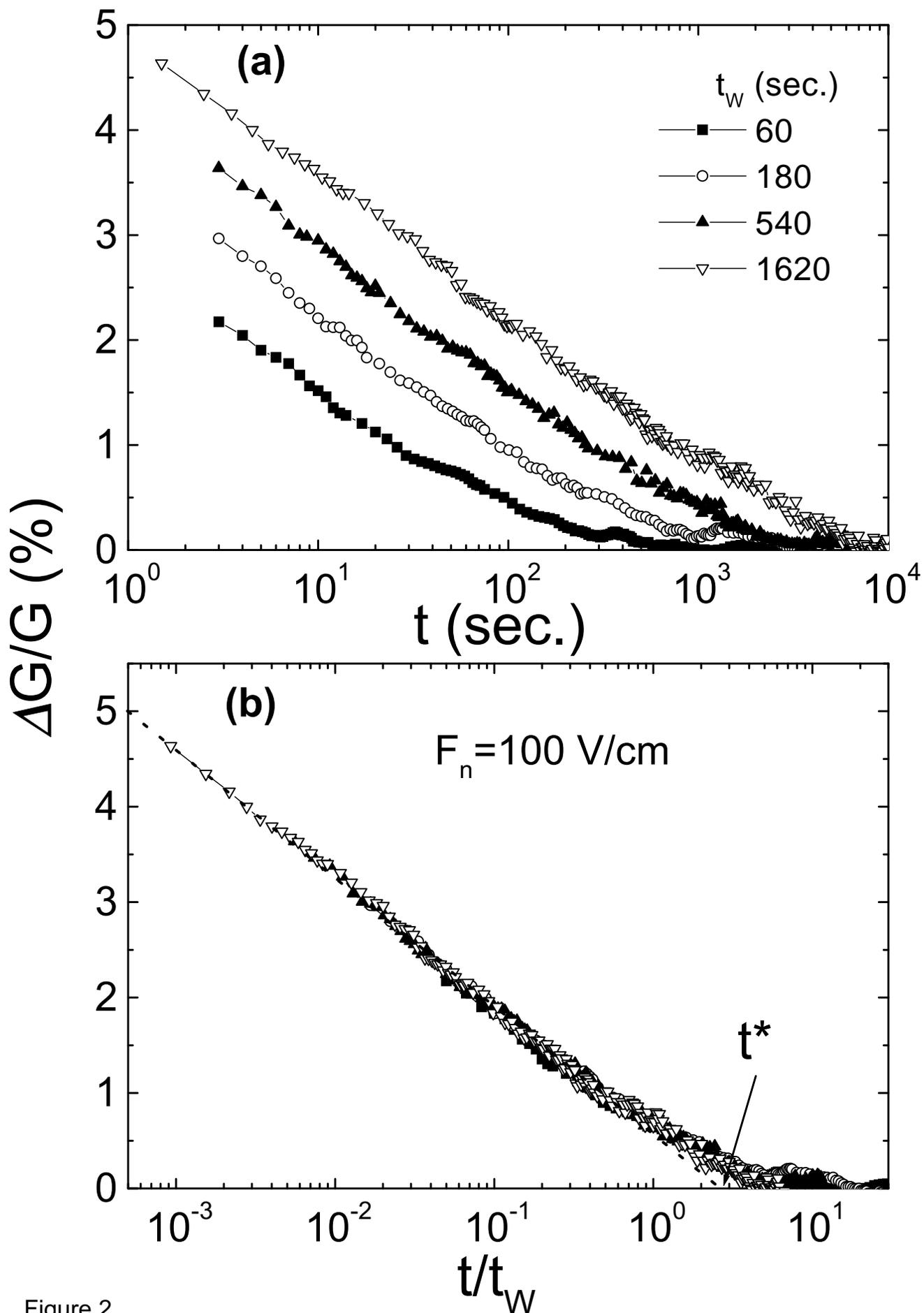
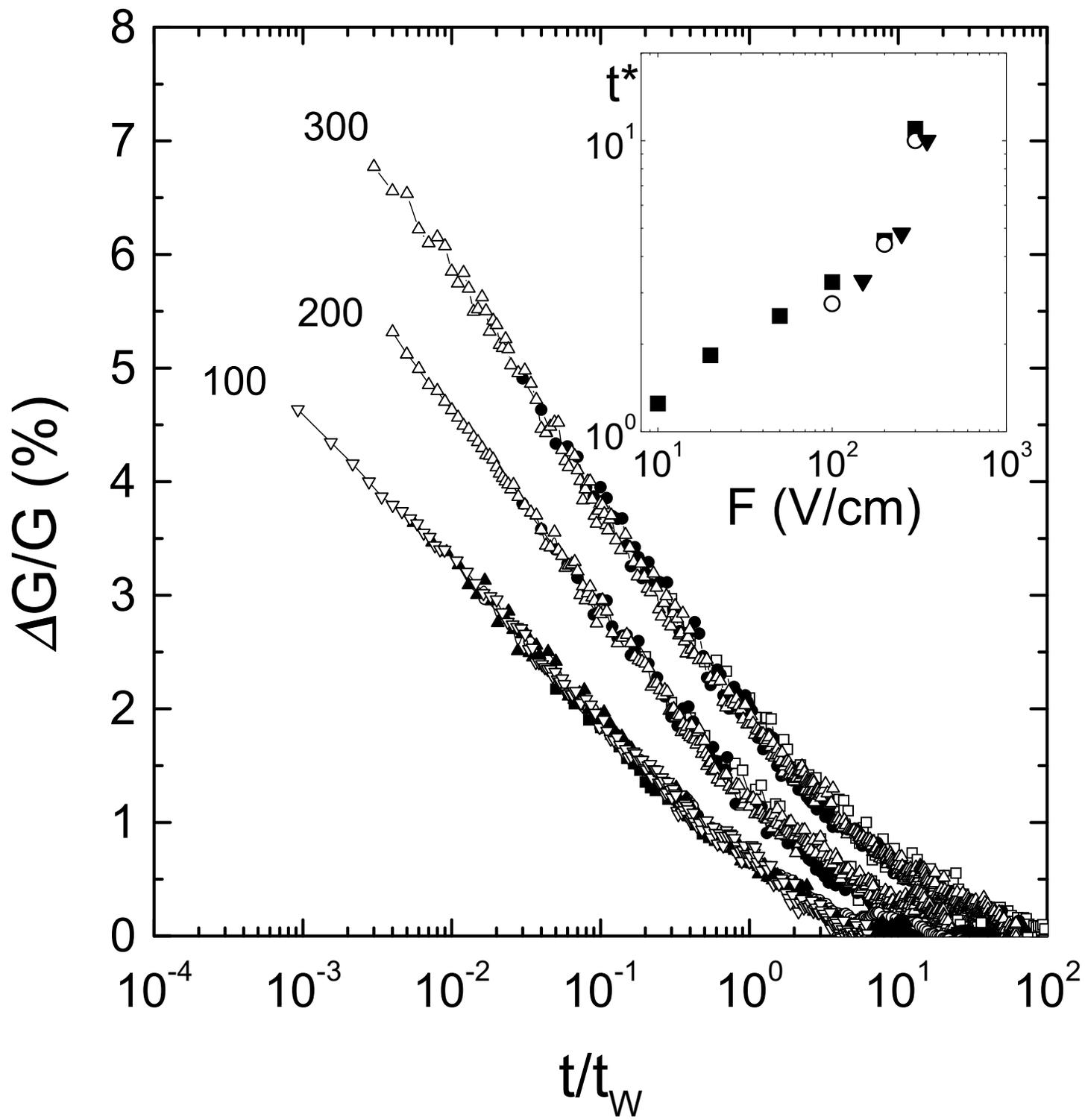
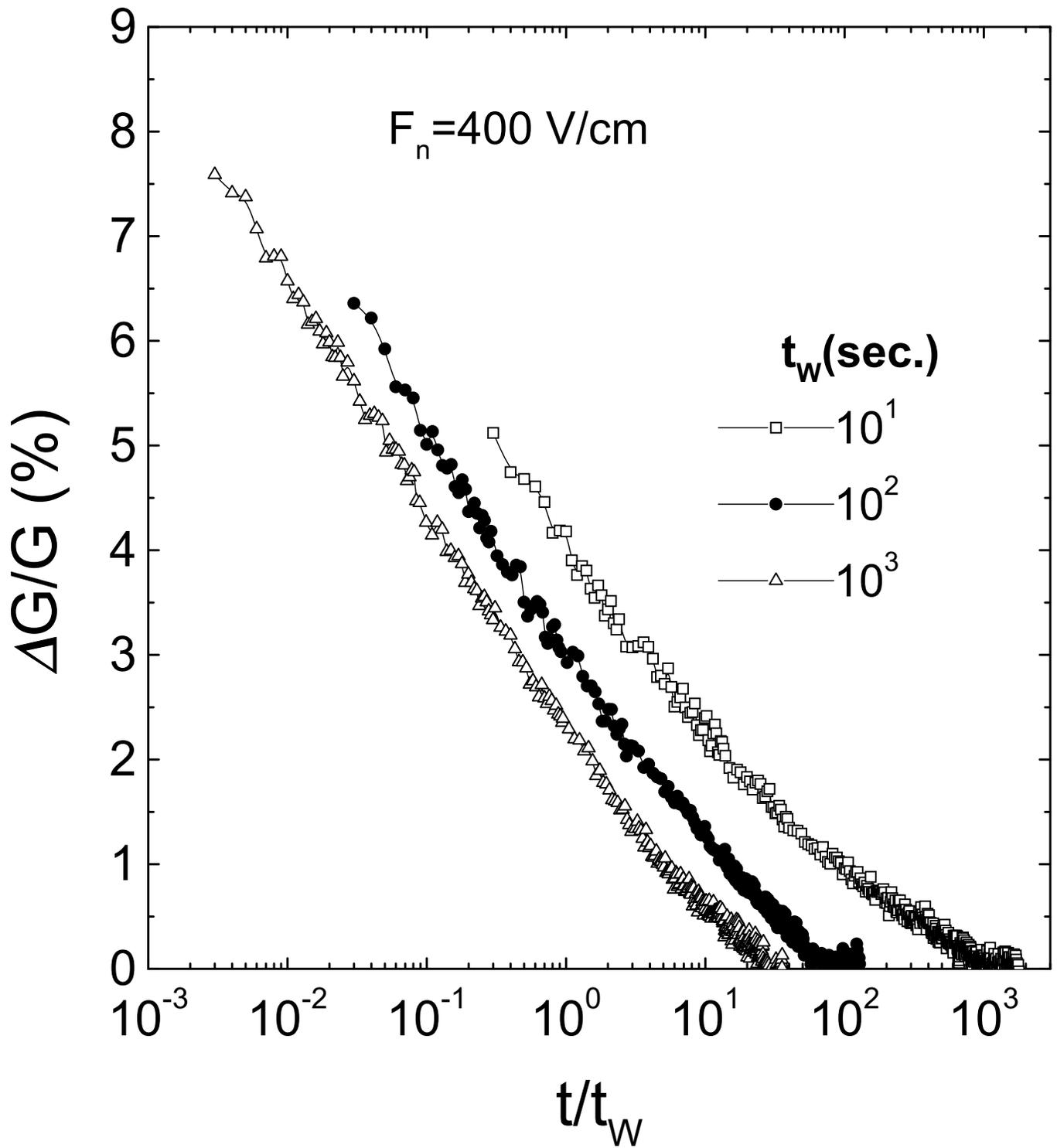
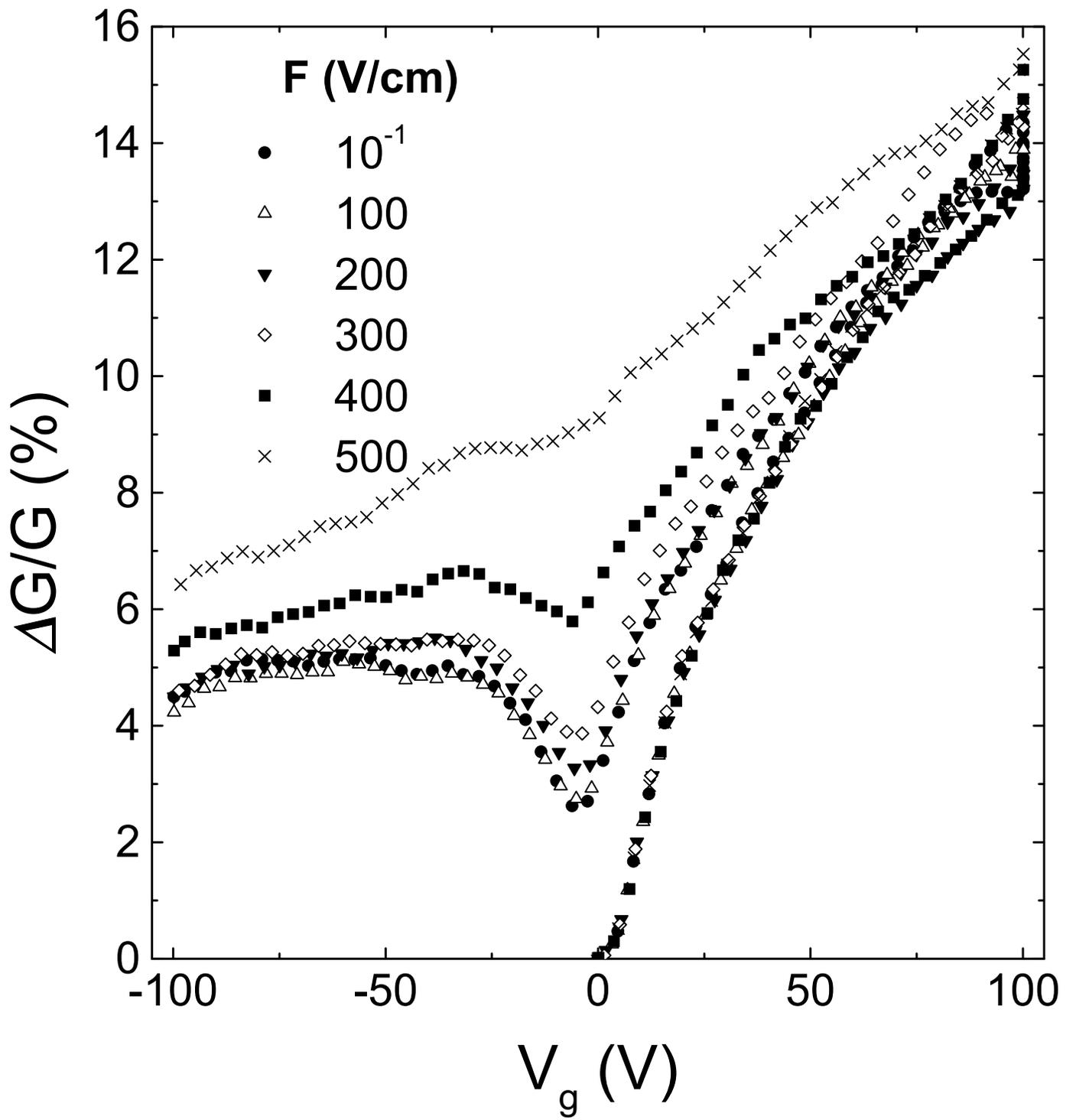
\end{document}